\documentstyle[preprint,aps]{revtex}
\begin{document}

\title{Stochastic Model for the Motion of a Particle on an Inclined
  Rough Plane and the Onset of Viscous Friction}

\author{G. G. Batrouni$^{1,2}$, S. Dippel$^{2}$, L. Samson$^{1}$}

\address{$^1$ ~GMCM, URA CNRS 804, Universit\'e de Rennes 1,
35042 Rennes Cedex, France}

\address{$^2$ ~ HLRZ, Forschungszentrum, D--52425 J{\"u}lich, Germany}

\maketitle
\begin{abstract}

  Experiments on the motion of a particle on an inclined rough plane
  have yielded some surprising results.  For example, it was found
  that the frictional force acting on the ball is viscous, {\it i.e.}
  proportional to the velocity rather than the expected square of the
  velocity. It was also found that, for a given inclination of the
  plane, the velocity of the ball scales as a power of its radius. We
  present here a one dimensional stochastic model based on the
  microscopic equations of motion of the ball, which exhibits the same
  behaviour as the experiments. This model yields a mechanism for the
  origins of the viscous friction force and the scaling of the
  velocity with the radius. It also reproduces other aspects of the
  phase diagram of the motion which we will discuss.

\end{abstract}

\vfill\eject
\bigskip
\bigskip

\section{Introduction}

A mixture of particles with different sizes can segregate under a wide
range of conditions and due to various mechanisms (usually a
combination) such as the ``Brazil nut effect''\cite{brazil},
shear\cite{shear}, percolation\cite{percolation},
convection\cite{knight}, or surface flow\cite{surf-flow}. Clearly the
problems of mixing and segregation have important consequences for
many industries. To understand this complicated phenomenon, it is
easiest to study the effects of the various mechanisms separately, if
possible.

In this paper we are interested in segregation due to flow which can
be seen, for example, in sand piles. When grains with various sizes
flow on the surface, one observes that the largest grains find their
way to the bottom of the pile while the smaller ones are stopped
farther uphill. This segregation is caused by the roughness of the
underlying surface on which the grains are flowing. In a much
simplified picture, one can consider the bulk of the sand pile as
providing a rough substrate on which the surface grains are flowing
and segregating. For experimental purposes, one can carry the
simplification farther and consider an inclined plane made rough by
sticking beads of a given radius, $r$, for example glass or sifted
sand, on contact paper and attaching it to the plane. This was done in
a series of experiments\cite{rennes1,rennes2,rennes3,rennes4} where
various properties of the motion of balls down this plane were
studied. We refer the reader to the references for details of the
experiments and results.  Here we will review some of the results
before proceeding to our theoretical model.

Three regimes were found\cite{rennes3,rennes4} for the motion of the
ball down the plane: A) A sticking or pinning regime where the ball
comes to a sudden stop after travelling some distance, B) a regime
where the ball attains a steady state with a constant (on average)
velocity independent of the initial release velocity, and C) a jumping
regime where the ball experiences big bounces and does not achieve a
steady state on the 2 meter long plane used in the experiments. One
expects that for a long enough plane, the ball will always reach a
steady state, but even so, the nature of the motion in this regime
appears to be different from that of region B.  However, because of
experimental difficulties, this regime remains mostly unstudied.

In region B, it was found that the constant (average) velocity
satisfies $\langle v\rangle \propto R^{\alpha}{\rm
  sin}(\theta)$\cite{scale}.  This is a surprising result because it
means that the force of friction acting on the ball is viscous, {\it
  i.e.} proportional to the
velocity\cite{rennes1,rennes2,rennes3,rennes4}. A straightforward
argument suggests that the friction force should be quadratic in the
velocity leading to a $\sqrt{{\rm sin}(\theta)}$ behaviour: It is
clear that the deceleration due to collisions is proportional to the
number of collisions per second times the velocity loss per collision.
The velocity loss per collision is proportional to the velocity
itself, and one can argue that the average number of collisions/second
is the velocity of the ball divided by the average distance between
surface beads. This gives a friction force which is quadratic in the
velocity, contrary to what is seen experimentally. Two assumptions
that would lead to viscous friction are 1) the velocity loss per
collision is a constant independent of the velocity, or 2) the number
of collisions per second is a constant independent of the velocity.
However, neither of these assumptions can be justified physically and
the explanation should be sought elsewhere.

It is tempting to draw an analogy between the motion of a ball down a
rough plane with peaks and troughs and the motion of particles in
random potentials\cite{ledoussal} using a Langevin equation to
describe the dynamics. However, such an approach cannot explain the
viscous dissipation because the viscous friction term is put in as an
assumption from the start.  In addition, even if one does assume the
viscous force and pursues this approach, the scaling of the velocity
with $R$, {\it i.e.} with the roughness which in this approach
characterizes the random potential, is not correctly reproduced. The
missing ingredient will be discussed in section III. A different
approach was taken in \cite{rennes2} where a Langevin equation, with
the viscous dissipation and the velocity scaling with $R$ built in,
was used to study other aspects of the motion such as the stopping
distance and its dependence on various parameters.

The approach we will take here is different still. We will consider
the one dimensional case, {\it i.e.} discs of radius $r$ are randomly
stuck on a line inclined at an angle $\theta$ with the horizontal, and
on which is released a disc of radius $R$. We then consider the
equations of motion of the disc, $R$, as it moves down the line, and
make as many simplifying assumptions as is reasonable, eventually
turning the deterministic equations into stochastic ones. Our goal in
this is not to have a detailed microscopic agreement with experiments:
That would be an unrealistic hope with a very simple model and is more
the domain of detailed numerical simulations. Rather, our goal is to
have a bare-bones stochastic model that agrees with experiments and
reproduces various properties of the motion such as the viscous
friction and the scaling of the velocity with $R$. With this model, we
succeeded in finding a simple mechanism for the viscous force and the
various scaling laws. In all this geometry plays a crucial role.

In section II we will construct the model clearly stating all our
assumptions, and in section III we will compare our results with the
experiments and make some comments. Section IV contains our
conclusions.

\bigskip
\bigskip

\section{The model}

Figure 1 shows the geometry of the system. Discs of radius $r$ are put
in a straight line, the distance between the surfaces of two
neighboring discs being $2\epsilon$, where $\epsilon$ is a random
number between $0$ and a maximum value, $\epsilon_M$. This line is
inclined an angle $\theta$ with respect to the horizontal, and we will
take the direction parallel (perpendicular) to it as the $x$ ($y$)
direction.  Therefore, the gravitational acceleration in the $x$ ($y$)
direction is $g{\rm sin}(\theta)$ ($-g{\rm cos}(\theta)$). The big
disc, radius $R$, moves from right to left with a velocity $\vec v$
and $x$ and $y$ components $v_x$ and $v_y$, making an angle $\phi$
with the line of discs, ${\rm sin}(\phi)=|v_y|/|\vec v|$.  As shown in
Fig.  1, $\gamma$ is the angle between the line perpendicular to the
line of discs, and the line connecting the centers of the big and
small discs at the point of contact. We will call this the angle of
contact which, it should be emphasized, is different from the angle of
incidence. Our sign convention is that when the collision is on that
side of the small disc which faces the approaching big disc (as in the
figure) $\gamma<0$, and when it is on the other side, $\gamma>0$.

Our first assumption is that we can ignore the rotation of the big
disc as it moves down the line. Clearly this assumption is not
justified if we are aiming at detailed microscopic comparison with
experiments. For example, neglecting rotation means that $v_t$, the
tangential velocity of the point of contact during a collision is due
entirely to the translational velocity of the disc. This is important
quantitatively because $\mu_t$, the tangential coefficient of
restitution, depends on $v_t$ (see below). However, molecular dynamics
(MD) simulations have shown\cite{sabine} that if we prevent the disc
from rotating as it bounces down the line and have only dynamic
friction acting as the tangential force, the qualitative features of
the motion, like the $\theta$ dependence of the velocity, do not
change.  Since we are interested here in the scaling properties of the
motion, we will neglect the rotation.  Therefore, with this
assumption, the only effect of the size of the disc is to contribute
geometrical constraints as we will see below.

Our second assumption is that after a collision, the velocity
components normal and tangential to the small disc at the point of
impact are related to the corresponding velocities before the
collision by
$$
|v_n^{\prime}| = \mu_n |v_n|,\eqno(1a)
$$
$$
v_t^{\prime} = \mu_t v_t,\eqno(1b)
$$
where $\mu_n$($\mu_t$) is the normal (tangential) coefficient of
restitution. There are absolute value signs in eq.(1a) but not eq.(1b)
because the coefficient of normal restitution is always positive while
that for tangential restitution can be negative (see eq.(9a,b)). The
angle of incidence, $a_i$ (not shown in the figure), is of course
given by ${\rm cos}(a_i)=|v_n|/|\vec v|$, and ${\rm
  sin}(a_i)=|v_t|/|\vec v|$, while the angle of reflection, $a_r$, is
given by ${\rm cos}(a_r)=|v_n^{\prime}|/|\vec v^{~\prime}|$, and ${\rm
  sin}(a_r)=|v_t^{\prime}|/|\vec v^{~\prime}|$, where the primes
denote values just after the collision.  Now consider the big disc
going down the inclined line colliding with the small discs, and let
us examine the $k$th collision. Let the $x$ and $y$ velocities just
before (after) the $k$th collision be $v_x(k)$ ($v_x^{\prime}(k)$) and
$v_y(k)$ ($v_y^{\prime}(k)$). It is easy to see from these definitions
and the simple plane geometry that
$$
v_n(k) = |v_y(k)|{\rm cos}(\gamma)-v_x(k){\rm sin}(\gamma),\eqno(2a)
$$
$$
v_t(k) = |v_y(k)|{\rm sin}(\gamma)+v_x(k){\rm cos}(\gamma),\eqno(2b)
$$
keeping in mind our convention for the sign of $\gamma$. The $x$
velocity after the $k$th collision is $v_x^{\prime}(k)=|\vec
v^{~\prime}|{\rm sin}(a_r+\gamma)$ which, after expanding the sine and
using the above definitions of the angle of reflection, gives
$$
v_x^{\prime}(k)=\mu_t v_t{\rm cos}(\gamma)+\mu_nv_n{\rm
  sin}(\gamma).\eqno(3)
$$
Combining this with eqs.(2a,b) allows us to express the $x$ velocity
just after the $k$th collision in terms of the $x$ and $y$ velocities
just before the collision as
$$
v_x^{\prime}(k)=v_x(k)\biggl (\mu_t{\rm cos}^2(\gamma)-\mu_n{\rm
  sin}^2(\gamma)\biggr )+|v_y(k)|\biggl (\mu_t+\mu_n\biggr ){\rm
  sin}(\gamma){\rm cos}(\gamma).\eqno(4a)
$$
A similar argument for $v_y(k)$ gives
$$
v_y^{\prime}(k)=-v_x(k)\biggl (\mu_t+\mu_n\biggr ){\rm
  sin}(\gamma){\rm cos}(\gamma)+|v_y(k)|\biggl (\mu_n{\rm
  cos}^2(\gamma)-\mu_t{\rm sin}^2(\gamma)\biggr ).\eqno(4b)
$$

Our next assumption, which is supported by MD
simulations\cite{sabine,sabine2}, is that the motion of the disc is
composed mainly of a series of small bounces with very little or no
rolling which we will, therefore, ignore. This means that the $k$th
collision will cause the disc to bounce and spend a time
$\delta\tau(k)$ in the air during which it will experience $x$
acceleration $g{\rm sin}(\theta)$. Therefore, its $x$ velocity just
before the $(k+1)$th collision is $v_x(k+1)=\delta\tau(k)g{\rm
  sin}(\theta)+v_x^{\prime}(k)$, which, combined with eq(4a), gives
$$
v_x(k+1)=\delta\tau(k)g{\rm sin}(\theta)+v_x(k)\biggl (\mu_t{\rm
  cos}^2(\gamma)-\mu_n{\rm sin}^2(\gamma)\biggr )+|v_y(k)|\biggl
(\mu_t+\mu_n\biggr ){\rm sin}(\gamma){\rm cos}(\gamma).\eqno(5a)
$$
What happens to the perpendicular velocity, $v_y$? If the disc, after
the collision shown in fig. 1, bounces up and lands on another small
disc but at the same $y$ value, its $v_y$ just before the new collision
is identical to that just after the previous collision. In general,
however, the $y$ value will be different because the big disc will
land on various different points of the small discs. However, since
we are typically interested in $R/r\ge 3$, we see that the variations
in $y$ are very small. We can, therefore, make the simplifying
assumption that $v_y(k+1)$, the $y$ velocity just {\it before} collision
$(k+1)$ equals $v_y^{\prime}(k)$, the $y$ velocity just {\it after}
collision $k$\cite{vy-assumption}. We then have
$$
v_y(k+1)=-v_x(k)\biggl (\mu_t+\mu_n\biggr ){\rm
  sin}(\gamma){\rm cos}(\gamma)+|v_y(k)|\biggl (\mu_n{\rm
  cos}^2(\gamma)-\mu_t{\rm sin}^2(\gamma)\biggr ).\eqno(5b)
$$

So, eqs.(5a,b) express the $x$ and $y$ velocities just before the
$(k+1)$th collision in terms of those just before the $k$th collision.
Although not written explicitly, to simplify the notation, it should
not be forgotten that the contact angle $\gamma$ appearing on the
right hand side of these equations is actually $\gamma(k)$, the angle
for the $k$th collision.  These equations are deterministic since,
given the initial conditions, one can, in principle, calculate for all
subsequent collisions the two terms that have not yet been specified,
$\delta\tau(k)$ and $\gamma(k)$.  Such a detailed microscopic approach
is not our goal in this paper. Instead, we want to transform
eqs.(5a,b) into stochastic processes by making $\gamma$ a stochastic
variable. To do that we need its distribution which must be based on
the underlying microscopics of the collisions. To simplify the
calculation of this distribution, we assume that the disc, $R$,
collides with a given small disc only once.  It is clear from fig. 1
that, for a given incident velocity, $\vec v$, or equivalently, for a
given angle $\phi$, the largest (positive) contact angle,
$\gamma_{max}(k)$, is obtained when the big disc collides with the
small disc while tangent to the line $L_2$. For this case,
$\gamma_{max}(k)=\phi(k)$. On the other hand, the smallest ({\it i.e.}
the largest negative) $\gamma_{min}(k)$ is obtained for the case shown
in the figure, with the big disc tangent to line $L_1$. Notice that
$L_1$ is tangent to the disc just to the right of the disc involved in
the collision. This shows how discs cast velocity dependent
``shadows'' on their neighbors thus restricting the area available for
a collision with a given bead. It is straightforward to show that
$$
{\rm sin}(\gamma_{min}(k))={\rm sin}(\phi(k)){\rm cos}(\alpha(k)) - {\rm
  cos}(\phi(k)){\rm sin}(\alpha(k)), \eqno(6a)
$$
where
$$
{\rm cos}(\alpha(k))=1 - 2{r+\epsilon \over
  r+R}{\sin}(\phi(k)),\eqno(6b)
$$
and where $2(r+\epsilon)$ is the distance between the center of the
disc under collision and that to its immediate right. Recall that
$\epsilon$ is a uniformly distributed random number between $0$ and
some maximum value $\epsilon_M$. This gives the range of the stochastic
variable, $\gamma(k)$, to be between a maximum value, $\phi(k)$ (where
${\rm sin}(\phi(k))=v_y(k)/|{\vec v(k)}|$), and a minimum value given
by eqs(6a,b). But we still need its distribution. To this end, we
assume that for a given collision, the point of disc $R$ that is
closest to line $L_1$ (in fig. 1 it actually touches it) is equally
likely to be anywhere between $L_1$ and $L_2$. The position of this
point determines the value of the contact angle $\gamma$, and
therefore knowing its probability distribution (which we assumed to be
uniform) gives the probability distribution for $\gamma$. Using
this, it is straightforward to show from the geometry that $\gamma$
is given by
$$
{\rm sin}(\gamma(k))={\rm sin}(\phi(k))\biggl ({R+y{\rm sin}(\phi(k))
  \over r+R}\biggr ) - {\rm cos}(\phi(k))\Biggl (1-\biggl (
{R+y{\rm sin}(\phi(k)) \over r+R}\biggr )^2\Biggr )^{1/2},\eqno(7a)
$$
where
$$
y={\rm uniform~ random~ number}\in \biggl [ {r \over {\rm
    sin}(\theta)}-2(r+\epsilon),{r \over {\rm
    sin}(\theta)}\biggr ].\eqno(7b)
$$

One sees that the distribution for $\gamma$ is a function of the
geometry ($R$, $r$, and $\epsilon$) and the incoming velocities ($v_x$
and $v_y$) which determine how much of the surface disc area is
available for collision. The distribution given by eqs. (7a,b) is
shown in fig. 2.

Equations (5a,b) and (7a,b) define most of the stochastic model. We
still need to specify $\delta\tau(k)$, the time between collisions,
when energy is fed into the system.  Perhaps a temptingly simple
hypothesis is that $\delta\tau(k)={\bar d}/v^{\prime}_x(k)$ where
${\bar d}$ is the average distance between the centers of the line
discs.  However, as explained in section I, it is easy to see that
this gives a friction force proportional to $v_x^2$, and not
$v_x$.  An alternative argument is to say that for a
perfectly smooth surface, the time between bounces is given only by
$v_y$ ($\delta\tau(k)= 2v_y(k)/g{\rm cos}(\theta)$), and assume the
same holds for the rough plane. Integrating the stochastic equations
of motion with this assumption produces two types of solution
depending on the parameters used: Either the ball energy is
continuously dissipated until it comes to a complete stop, or the ball
makes bigger and bigger jumps accelerating all the time and never
reaching a stationary state.  Clearly neither of these two
possibilities for $\delta\tau(k)$ is realistic. Since the surface is
rough, {\it both} $v_x$ and $v_y$ play a role in determining
$\delta\tau(k)$. As we shall see below, the balance between $v_x$ and
$v_y$ plays a crucial role in determining the properties of motion.

To motivate our choice for $\delta\tau(k)$ note, from eqs. (4a,b),
that a collision with a negative $\gamma$ transfers velocity from the
$x$ to the $y$ direction, while the opposite happens for a positive
$\gamma$. Furthermore, for a negative $\gamma$ collision, the big disc
has to jump {\it over} the small disc in order to continue its motion,
while for a positive $\gamma$ collision the big disc can {\it always}
continue its motion. For these reasons, we assume that for a collision
with $\gamma$ negative (positive), $\delta\tau(k)$ is determined only
by $v_y$ ($v_x$). In other words, for $\gamma(k)<0$
$$
\delta\tau(k)={2v_y^{\prime}(k)\over g{\rm cos}(\theta)},\eqno(8a)
$$
and for $\gamma(k)\ge 0$
$$
\delta\tau(k)= {2(r+\epsilon) \over v_x^{\prime}(k)}.\eqno(8b)
$$ Note that in eq.(8b) we are assuming that the distance travelled to
the next collision is well approximated (on average) by
$2(r+\epsilon)$.  Of course in reality, $v_x$ and $v_y$ {\it together}
determine the time between collisions. This can be calculated but
would not reveal the underlying reasons for the viscous force which we
are trying to explain.  Equations (8a,b) give a simplification that
attempts to understand the separate roles of the parallel and
perpendicular velocities.

The last ingredient in our model is the experimental observation that
the tangential coefficient of restitution is strongly dependent on the
angle of incidence\cite{foerster}, given by the normal (tangential)
components of the velocity, $v_n$ ($v_t$) (see eqs.(2a,b)). It was
found in reference \cite{foerster} that the tangential coefficient of
restitution is well parameterized by
$$
\mu_t(k)= 1-3.5(1+\mu_n)\mu {v_n(k) \over v_t(k)},\eqno(9a)
$$
in collisions that involve gross sliding, whereas in collisions that
do not
$$
\mu_t(k)= -\beta.\eqno(9b)
$$
In these equations, $\mu$ is the coefficient of friction, and
$\beta$ is the tangential coefficient of restitution in the absence of
sliding. The normal and tangential velocities are given by eqs.(2a,b).
We mention here that one consequence of our ignoring the rotation of
the big disc is that the tangential velocity is given only in terms of
the translational velocities $v_x$ and $v_y$. However, rotation
modifies the tangential (but not the normal) velocity and therefore
the effective $\mu_t$ for each collision. This effective change in
$\mu_t$ will lead to a different average velocity, but not to a
qualitatively different behaviour. This is one justification for
ignoring rotation.

Now, our stochastic process is completely defined by eqs. (5a,b),
(7a,b), (8a,b) and (9a,b). This stochastic process is nonlinear and
cannot be solved exactly. However the equations can be easily iterated
numerically, and the statistical properties of the motion of the disc
down the line studied in detail. The initial conditions are random
choices for $v_x(1)$, $v_y(1)$ and $\gamma(1)$ from which everything
else follows simply by generating the random number $y$. If a
collision results in a negative value for $v_x$ we consider the disc
to have been pinned.

The equations defining our model can be further simplified by noting
that all the angles are very small allowing us to expand all
trigonometric functions to second order in $\gamma$. However, the
resulting equations will still be nonlinear and require a numerical
integration. The utility of such an expansion would be for approximate
solutions, e.g. mean field.

\section{Results}

In this section we compare the results of our model with the
experiments. The first test of the model is whether it can reproduce
the three experimentally found
phases\cite{rennes1,rennes2,rennes3,rennes4}, the pinning phase A, the
constant velocity phase, B, where $\langle v_x\rangle \propto {\rm
  sin}(\theta)$, and the ``jumping'' phase, C. Figure (3a) shows
$\langle v_x\rangle$ as a function of ${\rm sin}(\theta)$ for three
values of $R$, the radius of the big disc. We clearly see regions that
are linear in sin($\theta$) through which we have fitted straight
lines.  Although we will not attempt to match our results exactly with
experimental values, it is beneficial to see if the theoretical
results are in reasonable general agreement with experiments. The
linear regions stretch from about $4.5$ to $14.5$ degrees (the precise
values depend on $R$, and for a given $R$ they depend on the
coefficients of friction and normal restitution), and the values of
$\langle v_x\rangle$ are of the order of $5$ to $15$ cm/sec, all in
very good general agreement with the experiments. Also in agreement
with experiments, we found that the average velocity in region B does
not depend on the initial velocity with which the ball was released.
This is an important property of the motion in this region and does
not apply in region C.  When we scale the curves in fig. (3a) by a
factor of $R^{-0.49}$ we get fig. (3b) which shows excellent scaling
in the linear region, again in agreement with the experiments. The
exponent we found, $-0.49$ differs from the experimental
value\cite{rennes2,rennes3}, $-1.4$. Changing the parameters of the
calculation, {\it i.e.} the values of the coefficients of restitution
and friction, changes the exponents only a little and does not bring
it close enough to the experimental value.  We think that this
difference could be due to the fact that the calculation is done in
one dimension while the experiments are two dimensional. It might also
be that our approximations, while maintaining the qualitative features
of the physics, have changed this exponent. This insensitivity of the
exponent to the values of the coefficients of restitution and friction
explains the experimental result\cite{irene} that for a given value of
$\theta$, $R$, and $r$, the average velocity is practically
independent of the material of the plane or the ball\cite{ftnote}. In
other words, having fixed $\theta$, $R$, and $r$, balls of plastic,
glass and steel, whose coefficients of restitution and friction are
not very different, were found to have the same terminal velocity.
Furthermore, this terminal velocity did not change when the surface of
the plane was changed from glass beads, radius $r$, to sand with
average grain radius $r$.

To check how linear these plots are, we show in fig. (4) $\langle
v_x\rangle-v_{fit}$ where $v_{fit}$ is the value of the velocity given
by the straight line fit to each of the curves. We see that for
sin($\theta$) between about $0.1$ and $0.25$ the deviation is
extremely small and the linear fit is excellent. For sin($\theta$)
larger than about $0.25$ we see that the $\langle v_x\rangle$
increases faster than linearly. We take this as the definition of
region C, which for brevity we will call the ``jumping'' regime
because, as mentioned in the introduction, the motion is dominated by
large jumps as compared with the small bounces in region B.  We see
from fig. (4) that the largest disc, dashed line, enters the jumping
regime earlier and faster than the smaller ones.  This is in agreement
with experimental phase diagrams\cite{rennes2,rennes3} and with
numerical simulations\cite{rennes5,sabine}. Experiments have also
shown that as the ball aproaches the jumping regime, C, it exhibits
intermittent behaviour where it still has a constant average velocity
but it shows bursts of acceleration and deceleration. Figure (5) shows
a plot of distance travelled as a function of time for
sin$(\theta)=0.53$ from our model.  Bursts of acceleration and
deceleration are clearly seen.  This intermittency gets much more
dramatic as one gets closer to the jumping regime (increasing
$\theta$) before the motion gets completely destabilized. The origin
of a burst of acceleration is a particularly big bounce, due to high
speed and a large negative $\gamma$, where the disc spends a long time
in the air, gaining a lot of energy which it has difficulty losing.
Then comes another big shock that dissipates a lot of the energy, and
the disc regains it former slower velocity.  This mechanism can be
inferred from the sharpness of slope changes ({\it i.e.} velocity
changes) in fig. (5).

Figure (4) also shows that as $\theta$ is decreased, the average
velocity decreases and eventually exhibits a sublinear dependence
before it enters the pinning regime. To study the transition
between the pinning regime, A, and the constant velocity regime, B, we
release several discs for each value of $\theta$ and $R$ (keeping $r$
fixed) one after the other and with slightly different initial
velocities. We consider the system in the pinning regime when at least
half the discs are stopped. As is clear from fig. (6), the pinning
regime quickly shrinks ({\it i.e.} moves towards smaller $\theta$),
and then disappears as $R$ is increased. This agrees with the
experiments, although in our calculation region A disappears faster.

Experiments have indicated\cite{rennes3,jan} that $\langle v_x\rangle$
does not depend on the ratio $R/r$ alone. For example, Jan {\it et
  al}\cite{jan} showed that changing $R$ while keeping the ratio $R/r$
fixed (they kept $R=r$), $\langle v_x\rangle$ grows like $\sqrt{R}$.
Figure (7) shows $\langle v_x\rangle R^{-0.5}$ as a function of
sin$(\theta)$ for $R=3, 4.5, 6, 7.5, 9$mm and $R/r$ fixed at $6$. The
figure exhibits excellent data collapse with an exponent that agrees
with the experimental value. The same scaling is observed in MD
simulations\cite{sabine}.

So, we see that the simplified stochastic model presented here has
reproduced remarkably well all the features of the experimental
results. But what is the mechanism behind the linear dependence on
sin($\theta$) and the scaling with $R$? As mentioned in section II, if
one argues that the time between collisions, $\delta\tau$, is set by
the parallel velocity, $v_x$, one gets a very stable dependence of the
velocity on $\sqrt{{\rm sin}(\theta)}$ implying a friction force that
is proportional to the square of the velocity. The {\it faster} the
disc moves, the {\it shorter} the time interval during which energy is
fed into the system. If, on the other hand, one assumes, as in
ballistic flight, that $\delta\tau$ is determined only by the
perpendicular velocity, $v_y$, then one gets very unstable motion. The
{\it faster} the disc, the {\it longer} the time interval during which
energy is fed into the system and the more difficult it is to
dissipate it. This drives the instability. In our simple model, the
time of flight is given by both $v_x$ and $v_y$ depending on the side
of the line beads that is impacted.

For small inclination angles, our model gives $\delta\tau$ that is
dominated\cite{dominate} by $v_x$ as shown in fig. (8a) for
sin$(\theta)=0.03$. This gives a dependence on sin($\theta$) that is
very close to a square root, as is seen for the same angle in fig.
(3a,b). The reason for this is that at such small angles, $v_y$ is so
small that the motion is mostly parallel to the line, {\it i.e.} in
the $x$ direction, which as mentioned above, gives a $\sqrt{{\rm
    sin}(\theta)}$ dependence. For large values of $\theta$, the
motion is very bumpy, $v_y$ is quite large and it dominates the
contribution to $\delta\tau$, as is seen in fig. (8c) for
sin$(\theta)=0.4$. At this value, we see that $\langle v_x\rangle$
increases faster than linearly with sin$(\theta)$, fig. (3a,b), as the
system is headed for the jumping regime, C. In between the two
regions of sublinear and superlinear dependence on sin$(\theta)$,
there is a region of competition between the $v_x$ and $v_y$
contributions to $\delta\tau$ as is seen in fig. (8b) for
sin$(\theta)=0.2$, which is in the heart of the linear regime. It is
this competition between the destabilizing influence of $v_y$ and the
strong stabilizing influence of $v_x$ that gives rise to the effective
viscous friction and linear dependence on sin$(\theta)$.

What about the scaling observed in figs. (3b) and (7)? We found that
if we use a tangential coefficient of restitution, $\mu_t$, that is
constant instead of as in eq. (9a,b), the scaling exponent for fig.
(3a) is $0$, {\it i.e.} all the curves always overlap completely in
the linear regime. Furthermore, we found that with this assumption, it
is the {\it smaller} discs that enter the jumping regime, C, first, in
clear contradiction with the experiments and intuition. Therefore the
experimentally based eqs. (9a,b) are crucial to obtaining the
correct behaviour. On the other hand, we found that taking $\mu_t$ to
be constant does not change the behaviour in fig. (7) ({\it i.e.} when
$R/r=$constant) by much: the exponent is changed from $0.5$ to $0.43$
and the scaling is still clearly visible. This interesting result
indicates that although, in our model, the exponents for figs. (3b)
and (7) are very similar, the two scaling effects appear to have
different origins!

It is worth emphasizing that geometry is playing a dominant role in
this model. The distribution of the stochastic variable, $\gamma$,
given by eqs. (7a,b), is dominated by the velocity dependent shadows
cast by discs on their neighbors, constraining the cross section
available for collisions. In addition, this distribution is clearly
crucial in deciding which of eqs. (8a) or (8b) determines the time of
flight whose role is discussed above. Because of all this, we see that
the allowed collisions and the velocities are strongly dependent on
the geometry which also means that, because of eq. (9a), $\mu_t$ is
also dependent on the geometry.

\section{Conclusions}

We have presented in this paper a stochastic model for the motion of a
disc down an inclined line of smaller discs, separated by random
distances. The stochastic equations are based on the original
deterministic equations of motion to which we have added some
simplifying assumptions. Our two main assumptions are that we can
neglect rotation (supported by MD\cite{sabine}), and
that the motion of the disc is a series of small bounces, with one
bounce per line disc (supported by recent experimental results on the
two dimensional plane\cite{henrique} and by MD
simulations\cite{sabine}). This simple model, where geometry plays a
very important role, accurately reproduces the features of the
experimental data, and explains the origin of the viscous friction
force. This is seen to be the result of competition between ballistic
motion (with large $v_y$ determining the the time of flight) and
motion parallel to the plane (where $v_x$ determines the time of
flight). We also emphasized the importance of the dependence of the
tangential coefficient of restitution on the angle of incidence.

The next phase is to compare the results from this model with
experiments currently being done on the stopping distance of the balls
in regions A and B. The properties of the stopping distance in region
B have very important industrial applications in the segregation of
grains with different sizes.  We will also compare with experiments
the results of our model for the longitudinal dispersion of the
velocity and its dependence on the sizes of the balls and the
inclination angle. For the longer term, we would like to generalize
the model to a two dimensional plane, where we can make more
quantitative comparisons with experiments including results for the
transverse motion which is absent on a one dimensional line.

\acknowledgements We wish to thank D. Bideau, Ph. de Forcrand, A.
Hansen, C. Henrique, I. Ippolito, J. Sch\"afer, K. Taylor, and D. Wolf
for very valuable discussions. G.G.B. and L.S. acknowledge support
from the GdR CNRS ``Physique des Milieux H\' et\' erog\` enes
Complexes''.

\vfill\eject

\centerline{\bf FIGURE CAPTIONS}

\begin{itemize}
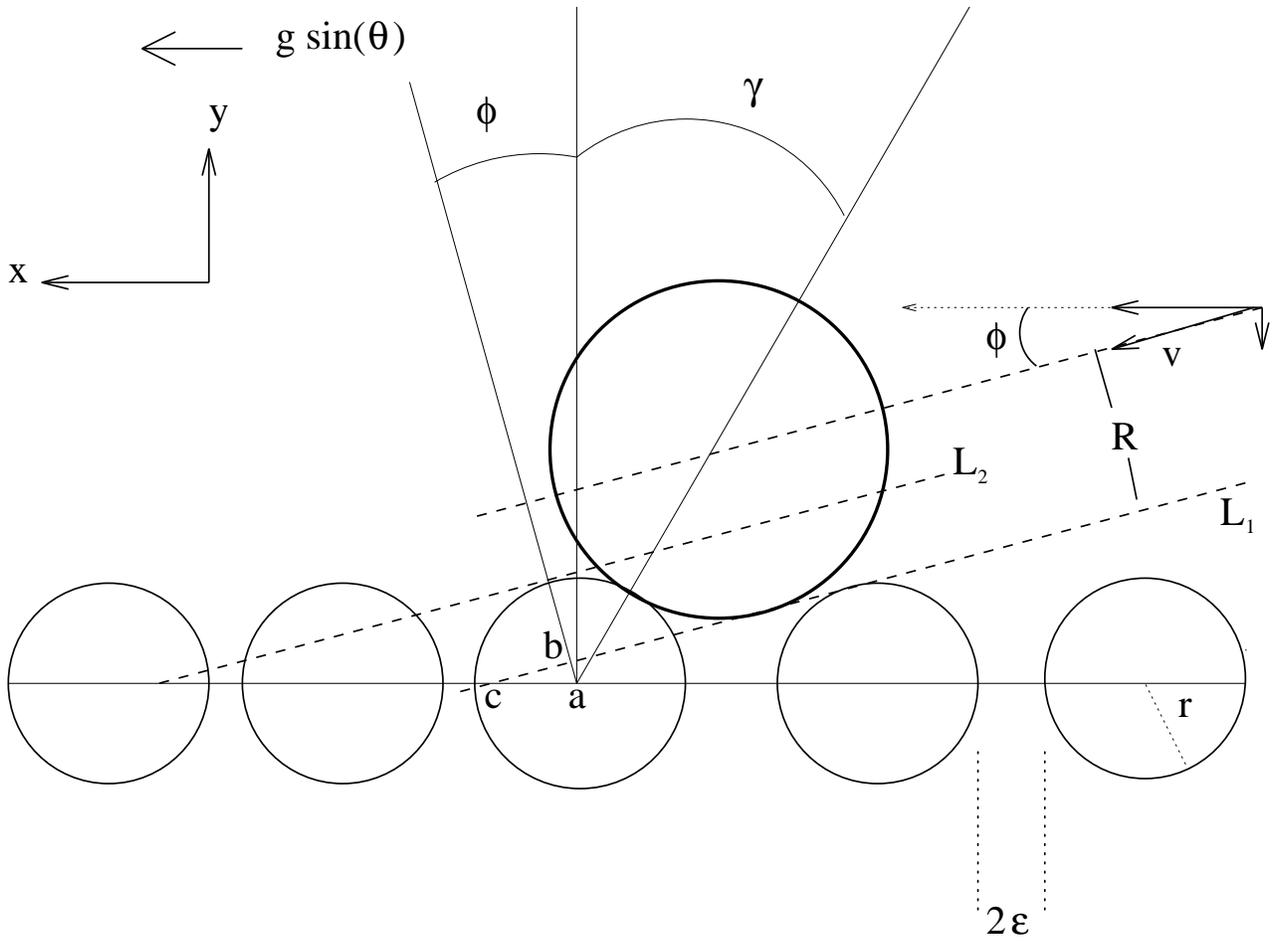
\item[FIG. 1] Shows the geometry of the inclined line of discs, $r$,
  on which we release a disc $R$. The collision depcited is that for
  the largest negative value of $\gamma$, the contact angle.

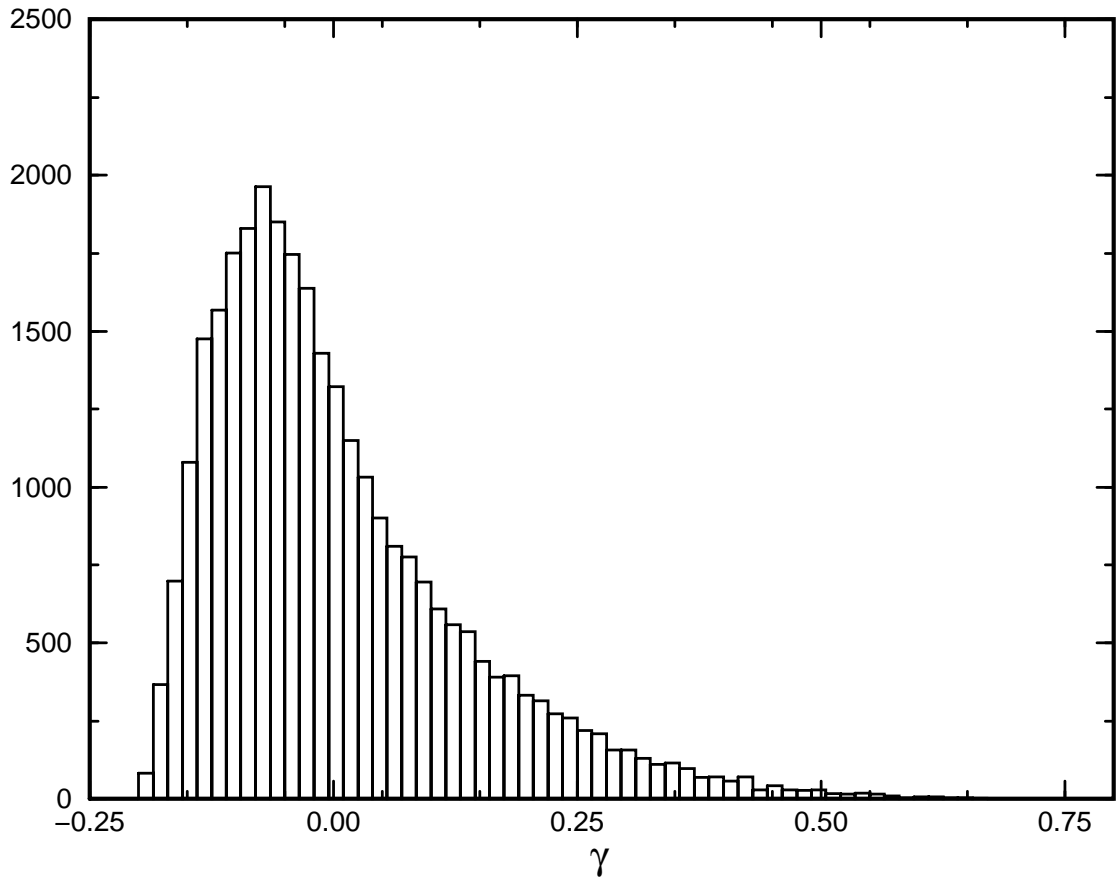
\item[FIG. 2] A histogram showing the distribution of the stochastic
  variable $\gamma$ (in radians) given by eqs. (7a,b). Recall that in
  our convention, $\gamma<0$ if the collision is as shown in fig. 1,
  and $\gamma>0$ if it is on the other side of the perpendicular.
  ${\rm sin}(\theta)=0.3$, $R=0.3$cm, $r=0.05$cm, $\epsilon=0.01$cm,
  $\mu_n=0.9$, $\mu=0.142$ and $\beta=0.45$. See eqs. (5a,b), (7a,b),
  (8a,b) and (9a,b) for definitions of the parameters.

\item[FIG. 3a] Shows the dependence of $\langle v_x\rangle$ (in
  cm/sec) on sin($\theta$). There is a clear linear region in
  agreement with experiments. $R=3, 4, 5$mm for $\Box,\bigtriangleup,
  \bigtriangledown$, $\mu=0.142$, $\mu_n=0.9$, $\beta=0.45$, $r=0.5$mm.

\item[FIG. 3b] Same as fig. (3a) but with the vertical axis scaled by
  $R^{-0.49}$.

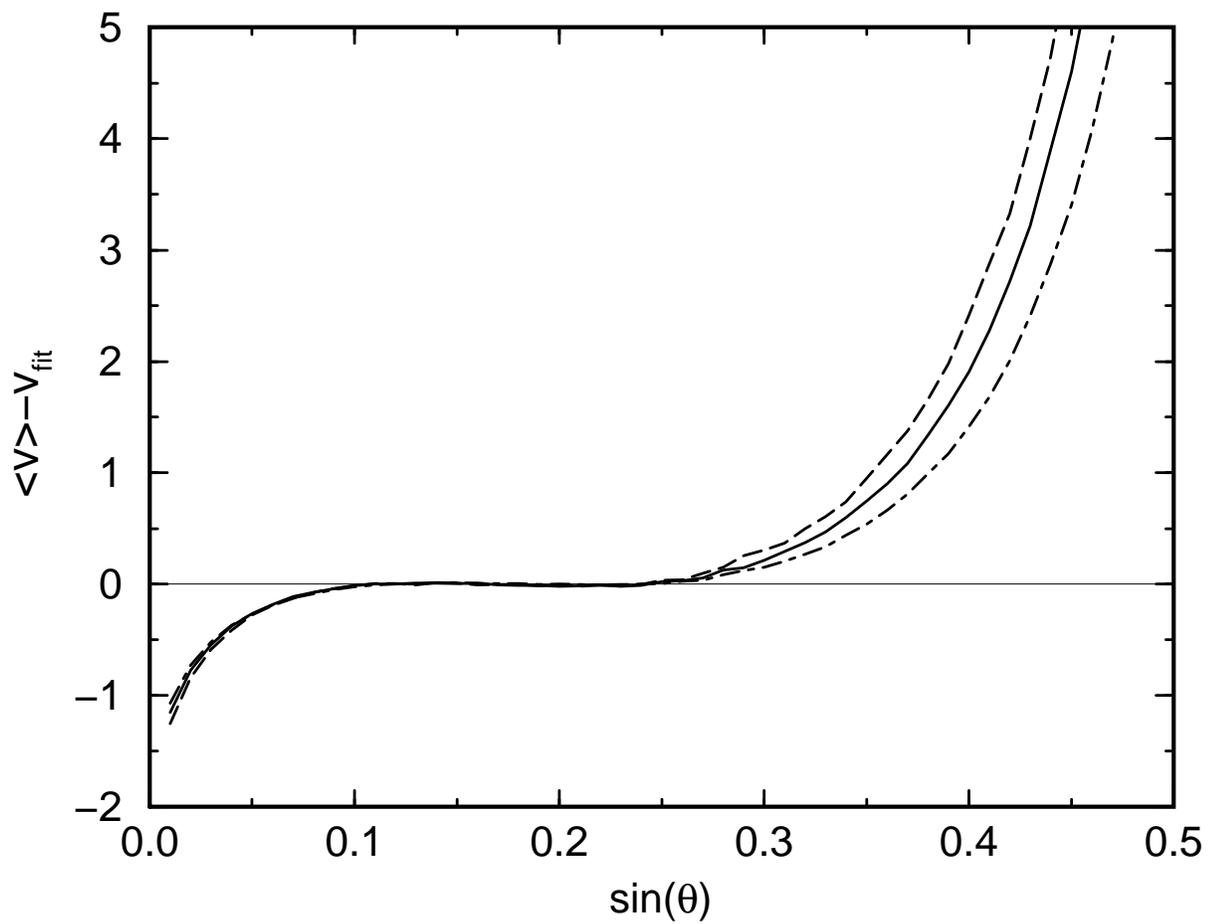
\item[FIG. 4] Deviations of $<v_x>$ from the straight line fits of
  fig. (3a). $R=3, 4, 5$mm for the dot-dash, solid, and dashed lines.

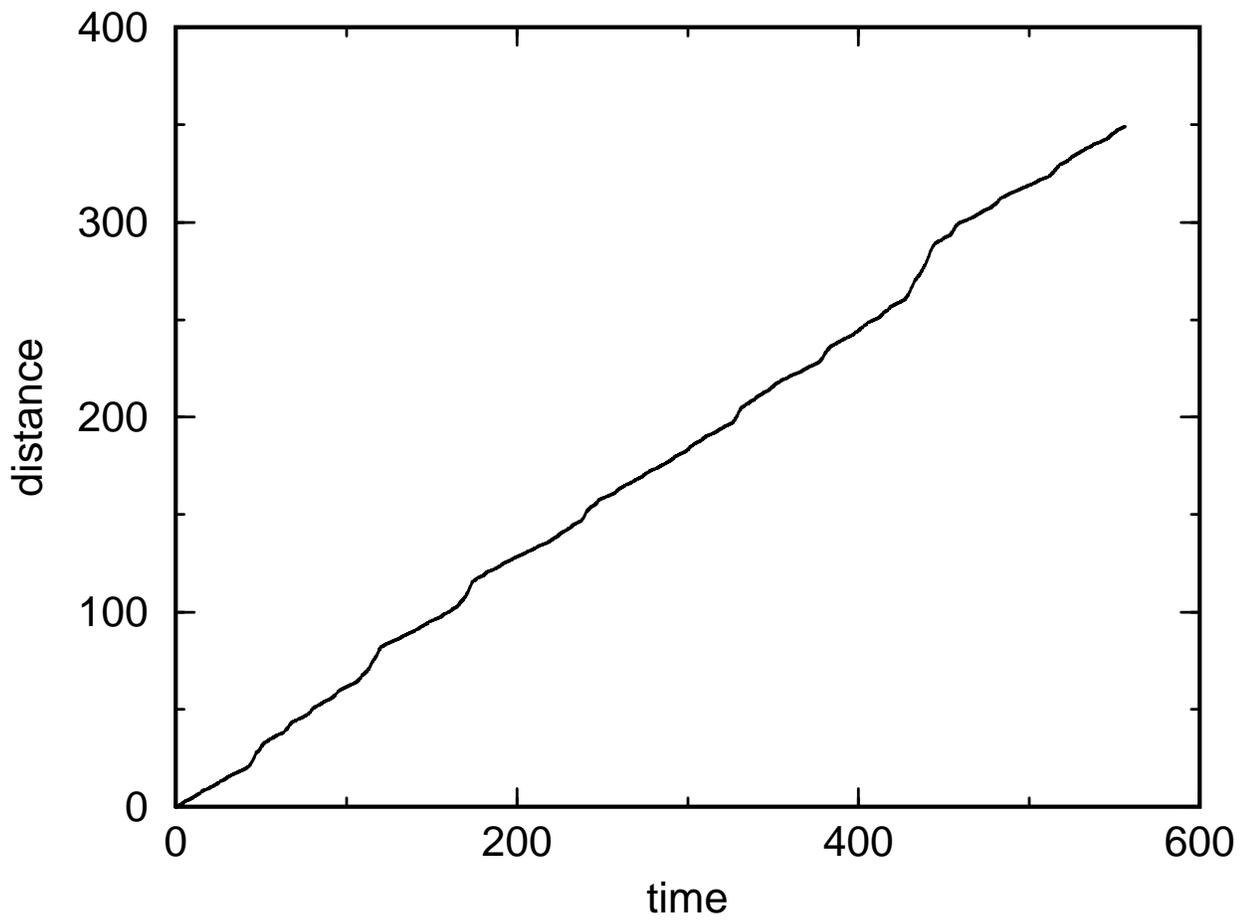
\item[FIG. 5] Distance travelled (in meters) as a function of time (in
  seconds) for a disc of radius $R=4$mm and an inclination angle
  sin$(\theta)=0.53$. All the other parameters are as in fig(3a,b).
  The slope gives the velocity parallel to the inclined line and
  exhibits bursts of acceleration and deceleration.

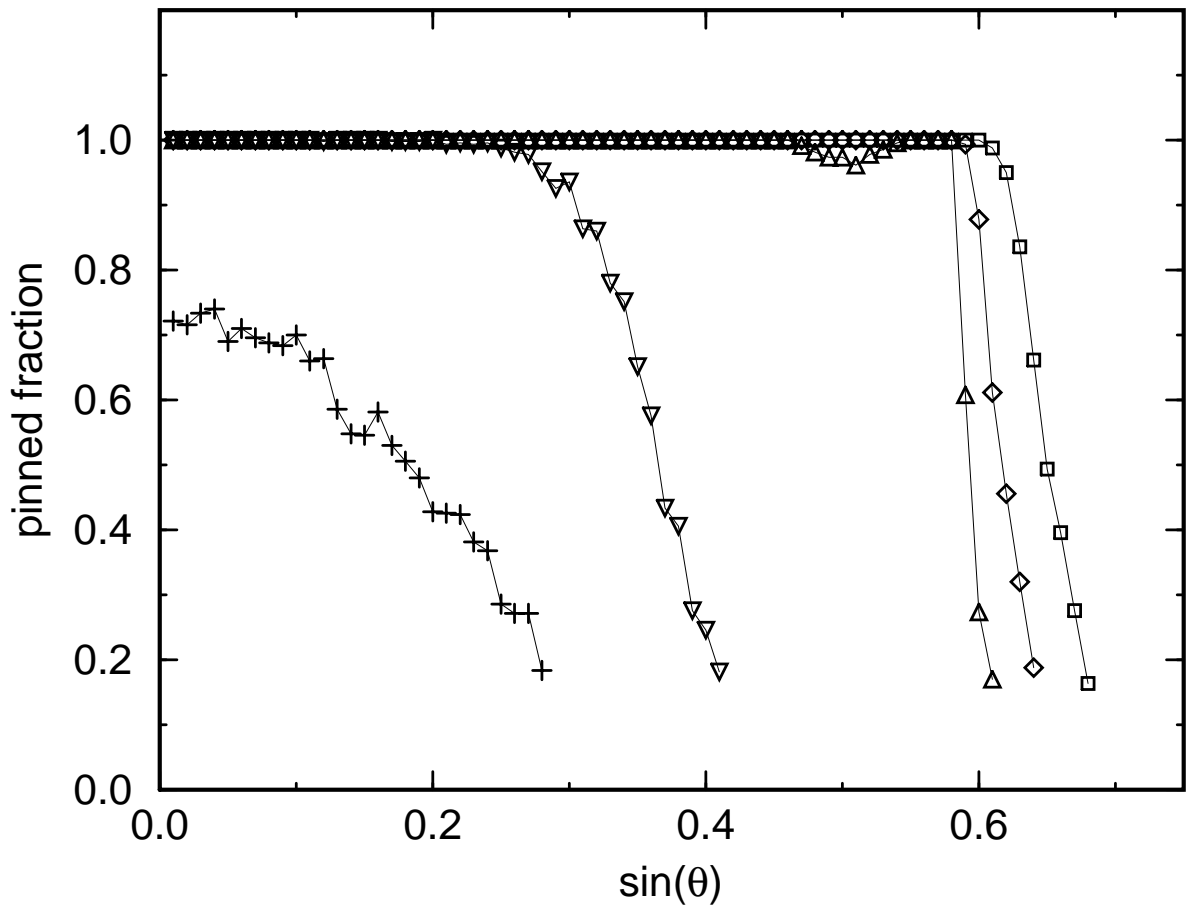
\item[FIG. 6] Fraction of pinned discs as a function of the
  inclination angle. $R=2$, $2.2$, $2.4$, $2.6$, $2.7$mm for $\Box,
  \diamondsuit, \bigtriangleup, \bigtriangledown, +$. The other
  parametrs are as in fig (3a,b). For each data point $500$ discs were
  released.

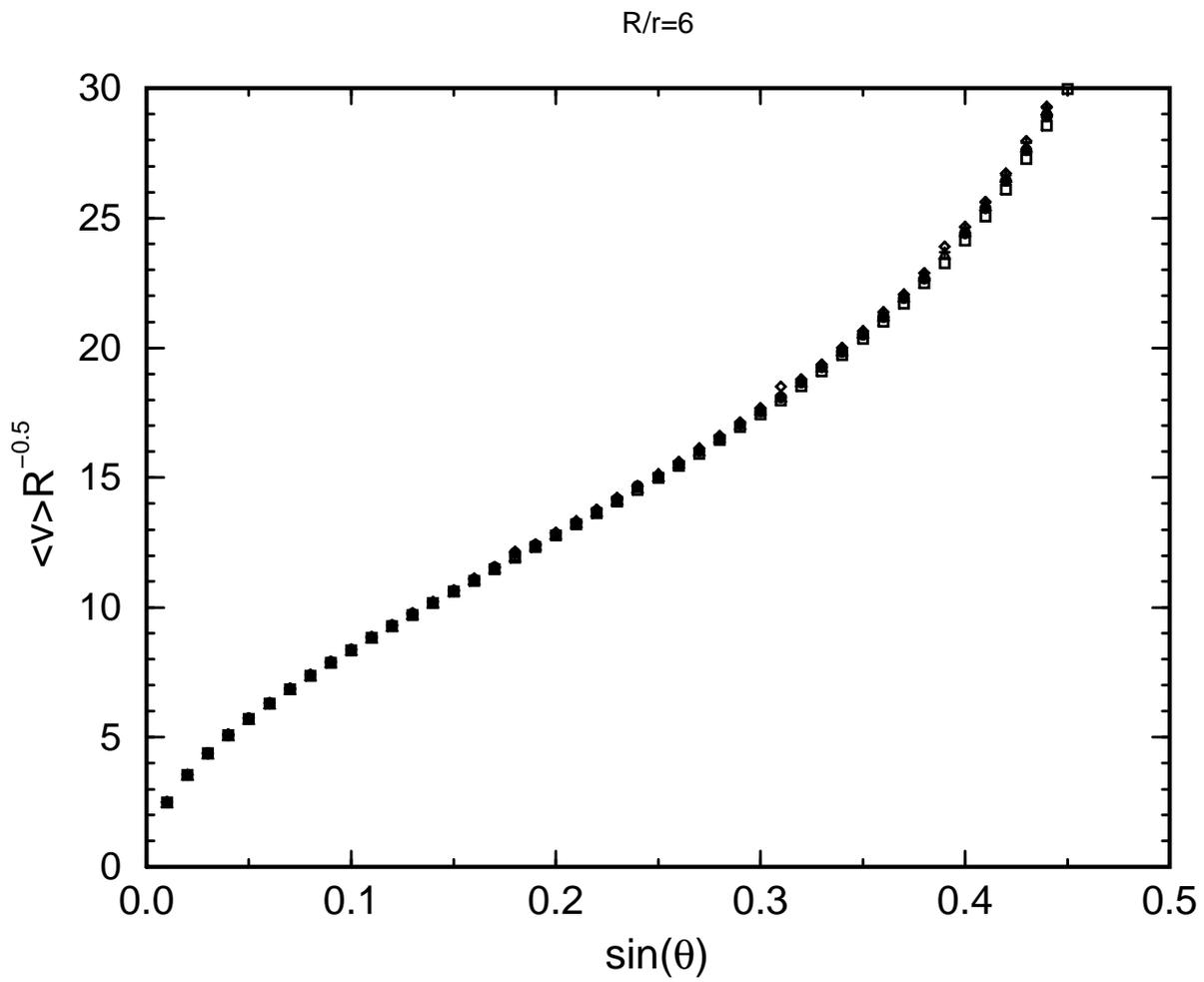
\item[FIG. 7] Scaling of $\langle v_x \rangle$ with $R$ when $R/r$ is
  kept constant. The data collapse is for $R=3, 4.5, 6, 7.5, 9$mm and
  the same parameters as in fig(3a,b).

\item[FIG. 8a] Histogram of times between bounces. The shaded
  histogram shows $\delta\tau$ for collisions at positive $\gamma$,
  {\it i.e.} where the time of flight is determined by $v_x$. The
  other histogram shows the case for negative $\gamma$, {\it i.e.}
  $\delta\tau$ is determined by $v_y$. Sin$(\theta)=0.03$, the other
  parameters are as in fig. (3a,b).

\item[FIG. 8b] As in fig. (8a) but with sin$(\theta)=0.2$.

\item[FIG. 8c] As in fig. (8a) but with sin$(\theta)=0.4$.

\end{itemize}


\begin{references}
\bibitem{scale} In references \cite{rennes1,rennes2,rennes3} the
  scaling is expressed as $(R/r)^{\alpha}$, but since the radius, $r$,
  of the beads on the plane is kept fixed, we will not use the ratio
  of the radii. Another reason for not doing that is that even with
  $R/r$ fixed, changing $R$ will change $\langle v\rangle$, as we will
  see later.
\bibitem{vy-assumption} We can include the effects of the fluctuations
  in $y$, but the model becomes more cumbersome and the results do not
  change.
\bibitem{ftnote} Clearly this can hold only within a certain range of
  parameters.
\bibitem{dominate} By ``dominate'' we mean that most of the energy fed
  into the system is done during bounces whose duration is determined
  by one of the velocities, $v_x$ or $v_y$.
\end{references}
\end{document}